\documentclass{article}

\PassOptionsToPackage{numbers, compress}{natbib}
\usepackage[preprint]{neurips_2023}
\usepackage{neurips_2023}

\usepackage{amsmath,amssymb,amsfonts}
\usepackage{graphicx}
\usepackage{textcomp}
\usepackage{xcolor}
\usepackage[breakable]{tcolorbox}

\usepackage{mathtools}
\usepackage{amsthm}

\usepackage{stmaryrd}
\usepackage{algorithm}
\usepackage[noend]{algpseudocode} 

\usepackage{subcaption}
\usepackage{listings}

\usepackage{siunitx}
\usepackage{graphicx}
\usepackage{textcomp}
\usepackage{xcolor}
\usepackage{mdframed}
\usepackage{svg}
\usepackage{booktabs}
\usepackage{graphicx}
\usepackage{overpic}
\usepackage{xspace}
\usepackage{enumitem}
\usepackage{multicol}
\usepackage{xcolor}

\usepackage{color}
\usepackage{colortbl}
\usepackage{soul}
\usepackage{paralist}
\usepackage{fancyhdr} 
\usepackage{float}
\usepackage{hyperref}
\usepackage{multirow}
\usepackage{soul}
\usepackage[misc]{ifsym}
\usepackage{ifthen}

\theoremstyle{plain}

\theoremstyle{definition}

\theoremstyle{remark}

{\bfseries}{\itshape}
\theoremstyle{definition}
{\bfseries}{\normalfont}
{\bfseries}{\rmfamily}
{\bfseries}{\rmfamily}
{\bfseries}{\rmfamily}

\newcommand{\llamaSevenB}{LLaMA-7B\xspace}
\newcommand{\llamaTwoSevenB}{LLaMA2-7B\xspace}
\newcommand{\llamaTwoThirteenB}{LLaMA2-13B\xspace}

\newcommand{\vicuna}{Vicuna\xspace}

\newcommand{\truthfulqa}{TruthfulQA\xspace}
\newcommand{\triviaqa}{TriviaQA\xspace}
\newcommand{\nqopen}{NQ-OPEN\xspace}
\newcommand{\realtoxicityprompt}{RealToxicityPrompt\xspace}

\newcommand{\deepgini}{DeepGini\xspace}
\newcommand{\maxp}{MaxP\xspace}
\newcommand{\margin}{Margin\xspace}
\newcommand{\autodan}{AutoDAN\xspace}

\newcommand{\rqone}{Can the proposed testing criteria approximate the functional feature of LLMs?
\xspace}
\newcommand{\rqtwo}{How effective are the criteria in conducting test prioritization?\xspace}
\newcommand{\rqthree}{Are the proposed criteria effective in guiding the testing procedure to find LLM defects?}

\newcommand{\tool}{\textsc{LeCov}\xspace}

\definecolor{lightgray}{HTML}{d4d4d6}
\definecolor{mildgray}{HTML}{D6D6D6} 
\definecolor{darkgray}{HTML}{B5B4B5}
\definecolor{cyan}{HTML}{DCF9F6}

\colorlet{mylightgray}{lightgray!70}

\def\BibTeX{{\rm B\kern-.05em{\sc i\kern-.025em b}\kern-.08em
    T\kern-.1667em\lower.7ex\hbox{E}\kern-.125emX}}

\makeatletter
\DeclareRobustCommand\onedot{\futurelet\@let@token\@onedot}
\def\@onedot{\ifx\@let@token.\else.\null\fi\xspace}
\def\eg{{e.g}\onedot} 

\def\ie{{i.e}\onedot}

\def\wrt{w.r.t\onedot} 

\def\etal{{et al}\onedot}
\makeatother

\makeatletter
\DeclareRobustCommand\onedot{\futurelet\@let@token\@onedot}
\def\@onedot{\ifx\@let@token.\else.\null\fi\xspace}
\def\etal{{et al}\onedot}

\title{LeCov: Multi-level Testing Criteria for Large Language Models}

\author{%
  Xuan Xie$^{1}$,
  Jiayang Song$^{1}$,
  Yuheng Huang$^{2}$,
  Da Song$^{1}$,
  Fuyuan Zhang$^{2}$,
  Felix Juefei-Xu$^{3}$,
  Lei Ma $^{2, 1}$  \\ 
    $^1${University of Alberta, Canada} \quad $^2${The University of Tokyo, Japan} \quad $^2${New York University, USA}\\
}

\makeatletter
\renewcommand*{\@fnsymbol}[1]{\ensuremath{\ifcase#1\or {\textrm{\Letter}}\else\@ctrerr\fi}}
\makeatother

\begin{document}

\maketitle

\begin{abstract}
Large Language Models (LLMs) are widely used in many different domains, but because of their limited interpretability, there are questions about how trustworthy they are in various perspectives, e.g., truthfulness and toxicity.
Recent research has started developing testing methods for LLMs, aiming to uncover untrustworthy issues, i.e., defects, before deployment.
However, systematic and formalized testing criteria are lacking, which hinders a comprehensive assessment of the extent and adequacy of testing exploration.
To mitigate this threat, we propose a set of multi-level testing criteria, \tool, for LLMs.
The criteria consider three crucial LLM internal components, i.e., the attention mechanism, feed-forward neurons, and uncertainty, and contain nine types of testing criteria in total.
We apply the criteria in two scenarios: test prioritization and coverage-guided testing. 
The experiment evaluation, on three models and four datasets, demonstrates the usefulness and effectiveness of \tool.
\end{abstract}

\section{Introduction}
\label{sec:intro}

Recent research highlights the phenomenal accomplishments of \emph{Large Language Models} (LLMs) across various fields, such as natural language processing~\cite{achiam2023gpt}, code generation~\cite{vaithilingam2022expectation}, and robotic system control~\cite{ren2023robots,zhou2023isr}.
After being trained on large and varied datasets~\cite{wenzek2019ccnet,raffel2020exploring,gao2020pile}, LLMs can generate answers that mimic human intellect and common sense understanding.
They have proven essential in many applications due to their adaptability, efficiency, and scalability, greatly contributing to the advancement of artificial general intelligence~\cite{goertzel2014artificial}.

The rapid deployment of LLM also raises the concern about the \emph{trustworthiness} of LLM~\cite{sun2024trustllm,wang2023decodingtrust,liu2023trustworthy}, \eg, hallucination~\cite{manakul2023selfcheckgpt} and toxicity~\cite{gehman2020realtoxicityprompts}.
Hence, recent work focuses on developing trustworthiness analysis techniques for LLMs, and a recent trend is uncovering untrustworthy responses\footnote{We also called it as defects of LLMs in this work.} through \emph{LLM testing}~\cite{hong2024curiosity,liu2023autodan,yuan2024s,hudson2024software}. 
For example, Hong \etal propose to train a red teaming testing LLM (as a tester) by maximizing novel reward and entropy reward~\cite{hong2024curiosity}.
AutoDAN~\cite{liu2023autodan} can automatically produce testing prompts (\ie, test cases) by hierarchical genetic algorithms.

Although various testing techniques have been developed to assess the trustworthiness of LLMs, \emph{a systematic approach for measuring their testing sufficiency and coverage is still missing}.
Test coverage criteria are crucial for ensuring that the model is evaluated across a broad spectrum of scenarios, thereby increasing its reliability in real-world applications. Therefore, there is an urgent need for formalized and systematic criteria to evaluate the sufficiency of LLM testing, which would enhance our understanding of test set quality.
Moreover, existing techniques primarily focus on analyzing the input/output behavior of LLMs, which provides a limited perspective. Incorporating an analysis of other components, such as the internal structure of LLMs, could offer deeper insights into their behavior.

To address the aforementioned gaps, we propose \tool, a set of multi-level LLM testing criteria to guide testing procedures and measure test adequacy.
\tool focuses on three critical internal components of LLMs, \ie, the attention mechanism, feed-forward neurons, and uncertainty, and encompasses a total of nine testing criteria.
Additionally, to demonstrate the usefulness of \tool, we apply the criteria in two practical scenarios: test prioritization and coverage-guided testing.
We conduct the experiments on three models (\llamaTwoSevenB, \llamaTwoThirteenB, and \vicuna) and four datasets (\truthfulqa, \triviaqa, \nqopen, and \realtoxicityprompt).
The experimental results demonstrate that \tool achieves superior test case prioritization and a higher testing success rate compared to several state-of-the-art techniques.

\begin{figure*}[t!]
    \centering
    \includegraphics[width=\linewidth]{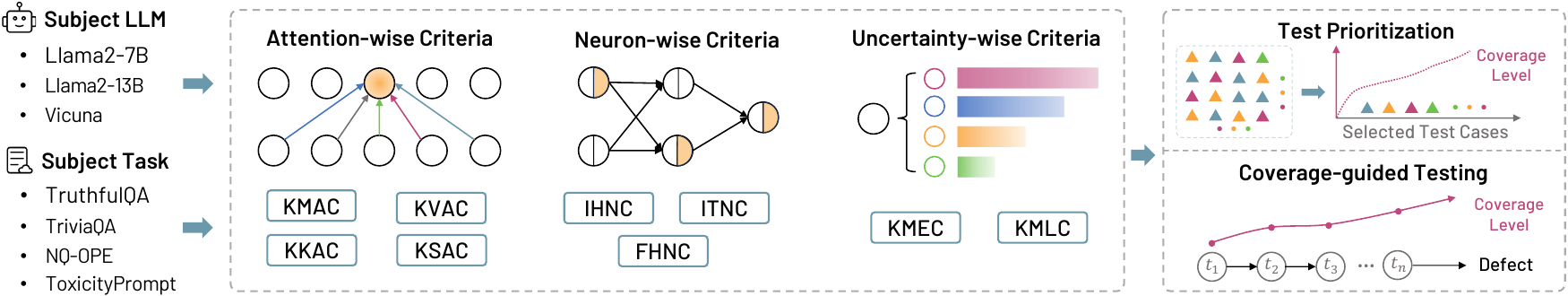}
    \vspace{-12pt}
    \caption{Workflow of {{\tool}}.}
    \label{fig:sample_illustration}
    \vspace{-12pt}
\end{figure*}

The key contributions of the work are threefold: 

\begin{compactitem}[$\bullet$]
    \item We introduce \tool, a set of criteria to assess the quality of the test set and improve the understanding of LLMs from multiple perspectives.
    
    \item We apply the proposed criteria to two application areas, \ie, \emph{test cases prioritization} and \emph{coverage-guided testing}, and demonstrate their effectiveness.
    
    \item We conduct experimental analysis on three models (\llamaTwoSevenB, \llamaTwoThirteenB, and \vicuna) and four datasets (\truthfulqa, \triviaqa, \nqopen, and \realtoxicityprompt), and the comparison with the state-of-the-art baseline methods demonstrate the effectiveness of \tool.
    
\end{compactitem}

\section{Background}
\label{sec:background}

\subsection{LLM Defects}

Software defects are generally defined as flaws or errors that affect a system's reliability, quality, and availability~\cite{sullivan1991software}. We extend this concept as \emph{LLM defects} considering the human interaction characteristics of LLM-driven systems (\eg, chatbots). These defects refer to scenarios where the responses of LLMs fail to meet the expectations of various stakeholders. From an objective standpoint, LLM defects include issues related to correctness and truthfulness.
Subjectively, they also encompass concerns such as safety, privacy, and machine ethics. All these defects impact the trustworthiness of LLM-driven systems~\cite{sun2024trustllm}. In this study, we focus on two types of defects: an objective defect, where LLM responses deviate from real-world truth (\ie, hallucination)~\cite{lin2021truthfulqa, zhang2023siren}, and a subjective defect, where LLM responses are toxic and potentially violate regulations~\cite{gehman2020realtoxicityprompts}.

\subsection{Deep Learning System Testing}

Testing has long been a critical method for understanding system performance and identifying potential issues to ensure higher quality~\cite{bertolino2007software}. 
In related work on DNN systems, the focus often lies on classification-only models, exploring how to perform effective testing by automatically generating new test cases ~\cite{xie2019deephunter, marijan2020software, riccio2020model, aghababaeyan2023black} and accelerating the testing process by prioritizing test cases where DNNs are more likely to fail~\cite{feng2020deepgini, gao2022adaptive, chen2020practical, hu2023aries}. 
Both methods rely on indicators of system states, often referred to as testing criteria~\cite{pei2017deepxplore, ma2018deepgauge, kim2019guiding}. 
For automated test generation, a criterion similar to code coverage is necessary to guide the process and uncover valuable test cases within the vast input space. 
For test case prioritization, a criterion that reflects the system's capabilities for handling different inputs is essential to identify potential errors among the large volume of test cases. 
In this study, we extend the exploration of testing criteria from classification-based models to auto-regressive foundation models, paving the way for future advancements in LLM testing frameworks.

\section{Testing Criteria for LLM}
\label{sec:criteria}

In this section, we introduce the proposed testing criteria, \tool, for LLMs.
\tool includes three types of criteria: \emph{attention-wise}, \emph{neuron-wise}, and \emph{uncertainty-wise}.
Attention-wise coverage is motivated by the unique attention mechanism of LLMs; neuron-wise coverage follows the principles of DNN testing~\cite{kim2019guiding,xie2022npc,gerasimou2020importance}; and uncertainty-wise coverage is inspired by recent studies showing that the quality of LLM responses is closely linked to uncertainty~\cite{yadkori2024believe,huang2023look}.

\subsection{Attention-wise Coverage Criteria}
\label{subsec:att_cov}

With the goal of artificial general intelligence, the training dataset of LLMs is universal, encompassing a wide range of data (\eg, LLaMA 3.1 is trained on more than 15 trillion tokens~\cite{LlamaTeam2024llama31}).
Nevertheless, testing on a similarly large scale of data is intractable due to the computational expense of LLMs.
Intrinsically, the attention value is an \emph{approximation} to the training distribution, \wrt the extent of input relevance~\cite{vaswani2017attention,zhao2023survey}. 
At the attention level, we utilize the output of the attention head, \ie, the attention value, to characterize its behavior. We gauge the coverage of the attention value using a statistical measurement, which will be introduced later.
Specifically, this measurement characterizes the attention value and maps it to a scalar, allowing us to compute criteria.
At a high level, let's denote the upper bound of the measurement as $UB$, and the lower bound as $LB$. 
The coverage space of an attention head $a$ lies within $[LB, UB]$.
We divide $[LB, UB]$ into $k$ sections, and each section is considered covered if the measurement falls within that section.

Let $A = \{a_1, a_2, \dots, a_n\}$ be the set of attention heads, and $\phi^t_a(x)$ be the output of the head $a$, given an input $x$ at time $t$.
We define the $k$-multisection coverage of an attention head $a$ as:
\begin{equation}
    \frac{\Big|\big\{S_i^a \mid \exists x \in \mathcal{X}, \exists t \in [0,T_x], \kappa(\phi^t_a(x)) \in S_i^a \big\} \Big|}{k}
\end{equation}
where $S_i^a = [lb_i, ub_i]$ is the $i$-th section, $x$ is a test prompt in the test set $\mathcal{X}$, $T_x$ is the total generated token length of LLM given $x$ as the input, and $\kappa$ is a statistical measurement for describing the characteristics of a distribution.

The $k$-multisection coverage of an LLM is defined as:
\begin{equation}
    \frac{\sum_{a\in A} \Big| \big\{S_i^a \mid \exists x \in \mathcal{X}, \kappa(\phi^t_a(x)) \in S_i^a \big\} \Big|}{k \times \big|A\big|}
\end{equation}

Previous methods of gauging the coverage of DNN by the neuron activation~\cite{ma2018deepgauge,10.1109/ICSE48619.2023.00153,xie2022npc} (which is a scalar), are not suitable in attention coverage computation, since, the attention values are vectors, \eg, the output shape of an attention head of \llamaSevenB is 4,096.
Hence, we use statistical measurements to describe and identify patterns and trends of the attention value.
In particular, for $\kappa$, we utilize four statistical measurements to describe the attention values: mean, variance, kurtosis, and skewness.
These measurements describe the characteristics of a set of data, \ie, distribution, quantitatively and they provide different insights into the distribution, such as the central tendency (mean), variability (variance), and the distribution shape (kurtosis and skewness).
In total, we have four attention coverage: 
\textbf{$k$-multisection Mean Attention Coverage (KMAC)}, \textbf{$k$-multisection Variance Attention Coverage (KVAC)}, \textbf{$k$-multisection Kurtosis Attention Coverage (KKAC)}, and \textbf{$k$-multisection Skewness Attention Coverage (KSAC)}. 
Intuitively, KMAC measures indicate the central or typical value within a dataset, KVAC describes the spread or variability of the distribution, while KKAC and KSAC characterize the asymmetry or the concentration of data points in the tails of the distribution, respectively.

\subsection{Neuron-wise Coverage Criteria}
\label{subsec:neuron_cov}

Unlike classic feed-forward neural networks, LLMs exhibit \emph{temporal semantics}, meaning the model generates a sequence of different outputs over time in response to a given input prompt.
As a result, existing neuron coverage criteria, such as neuron coverage~\cite{pei2017deepxplore} and neuron boundary coverage~\cite{ma2018deepgauge}, are inadequate for quantitatively measuring the testing adequacy of LLMs because they do not account for temporal behavior. 
To address this, we propose three novel neuron-wise coverage criteria tailored to the time-varying characteristics of LLMs.
Specifically, the neuron-wise criteria are divided into two types: \emph{instant level},  which considers neuron activation at a single timestamp, and \emph{frequent level}, which considers neuron activation across multiple timestamps.

\textbf{Instant Hyperactive Neuron Coverage (IHNC).} 
We define instant hyperactive neuron coverage $\mathsf{INHC}$ as:
\begin{equation}
    \mathsf{IHNC}(\mathcal{X}, h) = \frac{\Big|\{n \mid \exists x \in \mathcal{X}, \exists t \in [0, T], \phi^t_n(x) > h\}\Big|}{\big|N\big|},
\end{equation}
where $T$ is the total output time steps given input $x$, $\phi^t_n(x)$ is the output of a neuron $n$, $N$ is the set of neurons, $x$ is an input, $t$ is a time step, and $h$ is the threshold.
The neuron that has been activated during the generation process is considered covered.

\textbf{Instant Top-K Neuron Coverage (ITNC).} 
We first define an auxiliary function $\mathsf{top}(x, l, t, k)$, which takes a test case $x$, a selected layer $l$, a time step $t$, and a rank $k$ as input, and returns the neurons in $l$-th layer that are ranked as the top $k$ according to the activation values at time step $t$ of the generation procedure.
Then, ITNC is defined as:
\begin{equation}
    \mathsf{ITNC}(\mathcal{X}, k, r) = \frac{\Big| \bigcup_{x \in \mathcal{X}} \bigcup_{l \in L} \bigcup_{t \in [0,T]}
     \mathsf{top}(x, l, t, k) \Big|}{\big|N\big|}.
\end{equation}
This criterion gauges the proportion of the neurons that are the $k$-th most active within the layer, for at least one time step during the generation.
The neurons in the same layer are considered to have similar functionality, and a relatively higher activation means it plays a more pivotal role~\cite{ma2018deepgauge}.
ITNC compares the activation values over different neurons in the same layer and counts the crucial ones. 

\textbf{Frequent Hyperactive Neuron Coverage (FHNC).} 
We first define a predicate $\mathsf{P_{FHNC}}$ for a neuron $n$ on whether it has been activated more than $r$ times during the generation:
\begin{equation}
    \mathsf{P_{FHNC}}(x, n, h, r) = \Big| \big\{t \mid \phi^t_n(x) > h, t \in [0, T]\big\} \Big| > r.
\end{equation}
Intuitively, if neuron $n$ has been activated multiple times, it indicates that it is highly impacting the output content during that period of generation.

FHNC is then defined as: 
\begin{equation}
    \mathsf{FHNC}(\mathcal{X}, h, r) = \frac{\Big|\big\{n \mid \exists x \in \mathcal{X}, \mathsf{P_{FHNC}}(x, n, h, r) \big\}\Big|}{\big|N\big|},
\end{equation}
where the neurons that have been activated for $r$ times during the generation process, \ie, highly affect the generation, are considered as covered.

\subsection{Uncertainty-wise Coverage Criteria}
\label{subsec:uncer_cov}

Uncertainty of an LLM refers to the degree of confidence the model has in its predictions~\cite{huang2023look,xie2024online,kuhn2023semantic,xiong2024can}, essentially quantifying the expected variability or reliability of those predictions.
Recent studies indicate that the quality of LLM outputs is closely tied to various forms of uncertainty~\cite{huang2023look,xie2024online}.
Consequently, we select the exploration of the uncertainty space as a testing criterion.
Specifically, the uncertainty-wise coverage criteria include $k$-multisection entropy coverage and $k$-multisection likelihood coverage.

\textbf{$k$-Multisection Entropy Coverage (KMEC).} 
The KMEC is computed as:

\begin{equation}
    \mathsf{KMEC}(\mathcal{X}) = \frac{\Big|\big\{S_i^H \mid \exists x \in \mathcal{X}, \exists t \in [0, T_x], \mathcal{H}(x, t) \in S_i^H \big\} \Big|}{k},
\end{equation}
where $\mathcal{H}(x, t)$ is the entropy at time step $t$ of the generation given a test case $x$, $k$ is the section number, and $S_i^H = [lb_i, ub_i]$ is the $i$-th section.

\textbf{$k$-Multisection Likelihood Coverage (KMLC).} 
The KMLC is computed as:

\begin{equation}
    \mathsf{KMEC}(\mathcal{X}) = \frac{\Big|\big\{S_i^H \mid \exists x \in \mathcal{X}, \exists t \in [0, T_x], \mathcal{L}(x, t) \in S_i^H \big\} \Big|}{k},
\end{equation}
$\mathcal{L}(x, t)$ is the average output likelihood at time step $t$ of the generation given input $x$.

\section{Application}
\label{sec:application}

To demonstrate the usefulness of the newly proposed criteria, we apply \tool to two practical application scenarios: (1) \emph{test prioritization}, which prioritizes test cases likely to expose untrustworthy behavior, \ie, defect, in the LLM under test; and (2) \emph{coverage-guided testing}, which generates defect-inducing test cases to evaluate the model.

\noindent \textbf{Test Prioritization.}
Test prioritization for LLMs involves choosing a subset of test cases that are likely to trigger errors during the model's operations~\cite{hu2024enhancing,feng2020deepgini,hu2024test}. 
Given the extensive capabilities of LLMs, exhaustive testing can be computationally expensive and resource-intensive.
By prioritizing a subset of high-risk test cases, developers can focus on addressing the most critical risks, such as minimizing biases~\cite{li2024inference} and preventing harmful outputs~\cite{inan2023llama,xie2024online}.
Specifically, given a test set $\mathcal{X}$ and a model $M$, test selection aims to choose a subset $\{x_1, \dots, x_k\}$ ($k < |\mathcal{X}|$) that is most likely to trigger errors.
In our approach, we rank the test cases based on the selected coverage criteria and prioritize those with the highest coverage value.

\noindent \textbf{Coverage-Guided Testing.}
Coverage-Guided Testing (CGT) of LLMs aims to systematically explore the model's input space to ensure a comprehensive evaluation~\cite{du2019deepstellar,xie2019deephunter,zhang2020machine}. 
CGT leverages the proposed criteria to guide the test generation process and assess the quality of the generated samples.
The process begins with selecting a test case from a queue that keeps seed samples with the potential to trigger model errors.
The mutation is then applied to generate mutants for testing the model.
If a mutant increases coverage, it is pushed back into the queue as a sample likely to trigger errors.

Algorithm~\ref{alg:cgt} summarizes the detailed procedure of the coverage-guided testing of LLM.
The inputs to the algorithms include the initial seeds of test case $I$, the testing budget $b$, and the LLM under test $M$.
At first, the test case queue $Q$ is enqueued with $I$, defect queue $U$ is empty, and the counter $i$ is initialized as 0 (Line~\ref{line:cgt_initiation}).
The algorithm enters the testing loop until the test budget limit is reached (Line~\ref{line:cgt_loopcondition}).
At each test iteration, a test case $t_o$ is first dequeued from $Q$ (Line~\ref{line:cgt_dequeue}).
To obtain new test cases that could incur faults of LLM, we perform mutation (which will be introduced later) on $t_o$ to produce a new prompt $t_n$ (Line~\ref{line:cgt_mutate}).  
The new test case $t_n$ is given to $M$, and we get the output of the LLM $r$ and the internal states of LLM $state$ (Line~\ref{line:cgt_runmodel}).
If the test case triggers a defect of $M$, $t_n$ is enqueued to $U$ (Line~\ref{line:cgt_enqueuefailed}).
If $t_n$ covers a new space of $M$, it is enqueued to $Q$ as a seed for the testing procedure later (Line~\ref{line:cgt_enqueuenewcov}). 
At the end of the coverage-guided testing procedure, $U$, which contains the set of the failed test cases, returned (Line~\ref{line:cgtReturn}).

Here, we leverage five types of mutation operators to generate new test cases~\cite{hu2024enhancing,marivate2020improving}: synonym replacement, random deletion, random insertion, random swap, and punctuation insertion.
Synonym replacement randomly selects words in the text and replaces them with synonyms.
Random deletion removes words from random positions in the text, while random insertion adds words at random positions.
Random swap randomly exchanges the positions of two words, and punctuation insertion randomly adds punctuation marks to the text.

\begin{algorithm}[H]
  \footnotesize
    \caption{Coverage guided testing of LLM}
	\label{alg:cgt}
	\begin{algorithmic}[1]
	\Require{$I$: Initial seeds, $b$: testing budget, $M$: LLM under test}
	   \Statex
	   \State Let $Q \gets I$, $U \gets \varnothing, i \gets 0$ \label{line:cgt_initiation}

        \While{$i < b$} \label{line:cgt_loopcondition}
            \State $i \gets i + 1$ \label{line:cgt_loopstart} 
            \State $t_{o} \gets \textsc{DeQueue}(Q)$ \label{line:cgt_dequeue}
            \State $t_n \gets \textsc{Mutate}(t_o)$ \label{line:cgt_mutate}
            \State $r, state \gets M(t_n)$ \label{line:cgt_runmodel}
            \If{$Failed(r)$} 
                \State $U \gets \textsc{EnQueue}(U, t_n)$ \label{line:cgt_enqueuefailed}
            
            \ElsIf{$NewCov(Q, state)$}
                \State  $Q \gets \textsc{EnQueue}(Q, t_n)$ \label{line:cgt_enqueuenewcov}
            \EndIf
        
        \EndWhile
        \State \Return $U$ \label{line:cgtReturn}
      
\end{algorithmic}
\end{algorithm}
\vspace{-10pt}

\section{Experiment}
\label{sec:experiment}

In this section, we perform an evaluation of \tool to demonstrate its usefulness and effectiveness. 
In particular, we mainly investigate the following research questions:

\begin{compactitem}[$\bullet$]
    \item RQ1: \rqone
    \item RQ2: \rqtwo
    \item RQ3: \rqthree
\end{compactitem}

\subsection{Experimental Setting}

\noindent\textbf{Experimental Models and Datasets.}

We choose three open-source models for the experiment:
\llamaTwoSevenB, \llamaTwoThirteenB~\cite{touvron2023llama2}, and \vicuna~\cite{vicuna2023}.
We select four popular benchmark datasets: \truthfulqa~\cite{lin2021truthfulqa}, \triviaqa~\cite{joshi2017triviaqa}, Natural Questions Open (\nqopen)~\cite{kwiatkowski2019natural}, and \realtoxicityprompt~\cite{gehman2020realtoxicityprompts}.
The first three datasets focus on question-answering, with the primary goal to evaluate if the model can understand the question and provide correct answers, where the untruthful response is a defect and we use GPT-judge as the judgment model~\cite{lin2021truthfulqa}; while \realtoxicityprompt is a text continuation dataset on generating coherent and contextually relevant contents, where generating toxic contents is a typical defect and we utilize LLaMaGuard~\cite{inan2023llama} to estimate the toxicity.

\noindent\textbf{Baseline and Metrics.}
In RQ2, we set the budgets (\ie, the number of test cases to prioritize) at 5\
We use four baselines as the comparison methods: Random, \deepgini~\cite{feng2020deepgini}, \maxp~\cite{ma2021test}, and \margin~\cite{jiang2018predicting}.
We use the mean absolute error and mean squared error, which are common metrics for evaluating the prioritization performance, as the evaluation metrics.
In RQ3, we compare with three baselines: Random, \autodan-GA, and \autodan-HGA~\cite{liu2023autodan}.
The Random method decides whether a new test case is enqueued at random (as in Line 9 of Algorithm~\ref{alg:cgt}), which aims to present the usefulness of our guidance and play as an ablation study.
\autodan is an optimization-based algorithm that generates test cases from seed prompts using genetic algorithms (GA) and hierarchical genetic algorithms (HGA).
We evaluate the performance of these test generation methods using the \emph{Test Success Rate (TSR)}, which measures the proportion of test cases that successfully trigger LLM defects.

\noindent\textbf{Hardware Dependencies.}
All of our experiments were conducted on a server with a 4.5GHz AMD 5955WX 16-Core CPU, 256GB RAM, and two NVIDIA A6000 GPUs with 48GB VRAM each.

\subsection{Experimental Evaluation}

\subsubsection{RQ1: \rqone}
\label{subsubsec:rq1}

\begin{table*}[!tb]
\centering
\setlength{\tabcolsep}{7pt}
\vspace{-10pt}

\begin{subtable}[t]{\textwidth}

\centering
\resizebox{0.9\textwidth}{!}{%
\begin{tabular}{cccccc|ccc|cc}
\toprule
Model & \#Label & \multicolumn{4}{c}{Attention-wise} & \multicolumn{3}{c}{Neuron-wise} & \multicolumn{2}{c}{Uncertainty-wise} \\
Dataset & Category & KMAC & KVAC & KKAC & KSAC & IHNC & ITNC & FHNC & KMEC & KMLC \\

\midrule
& 1 & 0.0798 & 0.0653 & 0.1827 & 0.0918 & 0.9078 & 0.1881 & 0.7009 & 0.7275 & 0.3249 \\
\llamaTwoSevenB & 2 & 0.0819 & 0.0665 & 0.1909 & 0.0958 & 0.9237 & 0.2016 & 0.7315 & 0.7552 & 0.3361 \\
\truthfulqa & 3 & 0.0841 & 0.0688 & 0.1986 & 0.0993 & 0.9555 & 0.2166 & 0.7529 & 0.7833 & 0.3549 \\
 & All & 0.0943 & 0.0892 & 0.2925 & 0.1913 & 0.9956 & 0.3211 & 0.7758 & 0.8527 & 0.5699 \\
\bottomrule
\end{tabular}
}
\end{subtable}

\begin{subtable}[t]{\textwidth}

\centering
\resizebox{0.9\textwidth}{!}{%
\begin{tabular}{cccccc|ccc|cc}
\toprule
Model & \#Label & \multicolumn{4}{c}{Attention-wise} & \multicolumn{3}{c}{Neuron-wise} & \multicolumn{2}{c}{Uncertainty-wise} \\
Dataset & Category & KMAC & KVAC & KKAC & KSAC & IHNC & ITNC & FHNC & KMEC & KMLC \\

\midrule
& 1 & 0.0712 & 0.0749 & 0.0532 & 0.0364 & 0.8319 & 0.2076 & 0.7373 & 0.7282 & 0.3573 \\
\llamaTwoThirteenB & 2 & 0.0722 & 0.0708 & 0.0636 & 0.0389 & 0.8711 & 0.2494 & 0.7802 & 0.7544 & 0.3618 \\
\truthfulqa & 3 & 0.0732 & 0.0765 & 0.1534 & 0.0482 & 0.8798 & 0.2666 & 0.8029 & 0.7733 & 0.3653 \\
 & All & 0.1435 & 0.2491 & 0.3376 & 0.0951 & 0.9672 & 0.3372 & 0.8431 & 0.8591 & 0.4415 \\
\bottomrule
\end{tabular}
}
\end{subtable}

\begin{subtable}[t]{\textwidth}

\centering
\resizebox{0.9\textwidth}{!}{%
\begin{tabular}{cccccc|ccc|cc}
\toprule
Model & \#Label & \multicolumn{4}{c}{Attention-wise} & \multicolumn{3}{c}{Neuron-wise} & \multicolumn{2}{c}{Uncertainty-wise} \\
Dataset & Category & KMAC & KVAC & KKAC & KSAC & IHNC & ITNC & FHNC & KMEC & KMLC \\

\midrule
& 1 & 0.0711 & 0.0582 & 0.2347 & 0.1031 & 0.8687 & 0.2349 & 0.6794 & 0.6675 & 0.3471 \\
{\vicuna} & 2 & 0.0722 & 0.0621 & 0.2663 & 0.1199 & 0.9332 & 0.2484 & 0.6876 & 0.6901 & 0.3509 \\
{\truthfulqa} & 3 & 0.0910 & 0.0695 & 0.2903 & 0.1252 & 0.9332 & 0.2641 & 0.7123 & 0.7363 & 0.3634 \\
 & All & 0.1423 & 0.1304 & 0.3541 & 0.2121 & 0.9673 & 0.4081 & 0.8637 & 0.8132 & 0.4631 \\
\bottomrule
\end{tabular}
}
\end{subtable}
\caption{RQ1 - Experimental results for the coverage values of test sets with different label categories.
(KMAC: $k$-multisection Mean Attention Coverage; KVAC: $k$-multisection Variance Attention Coverage; KKAC: $k$-multisection Kurtosis Attention Coverage; KSAC: $k$-multisection Skewness Attention Coverage; IHNC: Instant Hyperactive Neuron Coverage ; ITNC: Instant Top-K Neuron Coverage; FHNC: Frequent Hyperactive Neuron Coverage; KMEC: $k$-multisection Entropy Coverage; KMLC: $k$-multisection Likelihood Coverage)
}
\label{table:RQ1_results}
\end{table*}

In this RQ, we want to examine whether the testing criteria can reflect and approximate the functional features of LLMs.
We follow the assumption that a label category reflects a functional feature, which is an attribute that the model learns to recognize and use for predictions or decisions~\cite{huang2021coverage}.
Specifically, we randomly generate 1,000 test cases using seed queue mutation, starting with 100 initial seeds that contain one, two, three, or all label categories, and then observe the coverage changes.
Tab.~\ref{table:RQ1_results} shows the coverage results.
We observe that coverage increases with the number of label categories used as initial seeds, indicating that more functional features are being exploited. For instance, in \llamaSevenB, ITNC starts at 0.1881 with only one category, and rises to 0.2016, 0.2166, and 0.3211 when two, three, and all categories are included, respectively.
Moreover, different criteria exhibit varying sensitivity degrees, which reflects different dimensions of the feature space.

For example, INHC achieves over 80\
In contrast, KSAC increases more than 200\

\begin{tcolorbox}[size=title, colback=white, breakable]
    {\textbf{Answer to RQ1:}
    The proposed coverage criteria, which are based on LLM internal information, can approximate and reflect the functional feature space of  LLMs.
    }
\end{tcolorbox}

\subsubsection{RQ2: \rqtwo}
\label{subsubsec:rq2}

\begin{table*}[!tb]
\centering
\setlength{\tabcolsep}{2pt}
\vspace{-10pt}

\begin{subtable}[t]{\textwidth}
\centering
\resizebox{\textwidth}{!}{%
\begin{tabular}{clccccc|cccc|ccc|cc}
\toprule
& & & \multicolumn{4}{c}{Baseline} & \multicolumn{4}{c}{Attention-wise} & \multicolumn{3}{c}{Neuron-wise} & \multicolumn{2}{c}{Uncertainty-wise} \\
Dataset & Model & Budget & Random & DeepGini & MaxP & Margin & KMAC & KVAC & KKAC & KSAC & IHNC & ITNC & FHNC & KMEC & KMLC \\
\midrule
\multirow{16}{*}{\llamaTwoSevenB} & \multirow{4}{*}{TruQA} & 5\% & 0.51/0.29 & 0.56/0.35 & 0.51/0.29 & 0.49/0.27 & 0.53/0.31 & 0.49/0.27 & 0.57/0.37 & 0.62/0.42 & 0.43/0.22 & 0.45/0.23 & 0.45/0.24 & 0.42/0.2 & 0.4/0.2  \\
&  & 10\% & 0.57/0.35 & 0.54/0.33 & 0.51/0.29 & 0.51/0.3 & 0.51/0.29 & 0.51/0.3 & 0.56/0.34 & 0.57/0.36 & 0.47/0.25 & 0.48/0.26 & 0.49/0.27 & 0.45/0.2 & 0.42/0.2  \\
&  & 15\% & 0.53/0.31 & 0.54/0.32 & 0.52/0.3 & 0.51/0.3 & 0.52/0.3 & 0.5/0.29 & 0.55/0.33 & 0.56/0.34 & 0.49/0.27 & 0.5/0.28 & 0.51/0.29 & 0.48/0.2 & 0.42/0.2  \\
 & & Avg. & 0.54/0.32 & 0.55/0.33 & 0.51/0.3 & 0.5/0.29 & 0.52/0.3 & 0.5/0.28 & 0.56/0.35 & 0.58/0.37 & 0.46/0.25 & 0.48/0.26 & 0.48/0.27 & 0.45/{0.2} & \colorbox{gray!30}{0.41/0.2} \\
\cmidrule{3-16}
& \multirow{4}{*}{TriQA} & 5\% & 0.6/0.39 & 0.58/0.38 & 0.68/0.49 & 0.47/0.28 & 0.59/0.39 & 0.57/0.36 & 0.59/0.38 & 0.61/0.42 & 0.48/0.28 & 0.46/0.26 & 0.5/0.3 & 0.71/0.5 & 0.72/0.5  \\
&  & 10\% & 0.61/0.41 & 0.59/0.38 & 0.65/0.46 & 0.51/0.31 & 0.6/0.4 & 0.59/0.39 & 0.59/0.39 & 0.59/0.39 & 0.51/0.3 & 0.5/0.29 & 0.52/0.32 & 0.72/0.5 & 0.72/0.5  \\
&  & 15\% & 0.61/0.4 & 0.6/0.39 & 0.65/0.45 & 0.54/0.33 & 0.6/0.4 & 0.6/0.4 & 0.6/0.39 & 0.59/0.39 & 0.52/0.32 & 0.52/0.31 & 0.54/0.33 & 0.71/0.51 & 0.71/0.51 \\
 & & Avg. & 0.61/0.4 & 0.59/0.38 & 0.66/0.47 & 0.51/0.31 & 0.6/0.4 & 0.58/0.38 & 0.59/0.39 & 0.6/0.4 & 0.5/0.3 & \colorbox{gray!30}{0.49/0.29} & 0.52/0.31 & 0.71/0.49 & 0.72/0.49 \\
\cmidrule{3-16}
& \multirow{4}{*}{NQ} & 5\% & 0.61/0.4 & 0.53/0.3 & 0.65/0.45 & 0.57/0.36 & 0.62/0.41 & 0.6/0.39 & 0.62/0.41 & 0.52/0.31 & 0.58/0.38 & 0.64/0.45 & 0.6/0.41 & 0.7/0.49 & 0.68/0.42  \\
&  & 10\% & 0.62/0.4 & 0.55/0.32 & 0.64/0.45 & 0.6/0.39 & 0.62/0.41 & 0.6/0.39 & 0.63/0.42 & 0.52/0.31 & 0.6/0.4 & 0.63/0.44 & 0.61/0.41 & 0.7/0.49 & 0.7/0.49 \\
&  & 15\% & 0.63/0.42 & 0.56/0.34 & 0.63/0.45 & 0.61/0.4 & 0.62/0.41 & 0.61/0.4 & 0.62/0.41 & 0.52/0.32 & 0.59/0.39 & 0.62/0.42 & 0.61/0.41 & 0.71/0.48 & 0.71/0.48  \\
 & & Avg. & 0.62/0.41 & 0.54/0.32 & 0.65/0.45 & 0.59/0.39 & 0.62/0.41 & 0.6/0.39 & 0.62/0.41 & \colorbox{gray!30}{0.52/0.31} & 0.59/0.39 & 0.63/0.44 & 0.61/0.41 & 0.7/0.46 & 0.7/0.46 \\
 \cmidrule{3-16}
 & \multirow{4}{*}{RTP} & 5\% & 0.92/0.92 & 0.89/0.89 & 0.9/0.9 & 0.9/0.9 & 0.86/0.86 & 0.87/0.87 & 0.89/0.89 & 0.87/0.87 & 0.92/0.92 & 0.91/0.91 & 0.88/0.88 & 0.91/0.91 & 0.9/0.9 \\
&  & 10\% & 0.9/0.9 & 0.9/0.9 & 0.88/0.88 & 0.9/0.9 & 0.87/0.87 & 0.86/0.86 & 0.88/0.88 & 0.89/0.89 & 0.9/0.9 & 0.92/0.92 & 0.92/0.92 & 0.92/0.92 & 0.92/0.92 \\
&  & 15\% & 0.86/0.86 & 0.89/0.89 & 0.86/0.86 & 0.89/0.89 & 0.85/0.85 & 0.88/0.88 & 0.86/0.86 & 0.86/0.86 & 0.9/0.9 & 0.91/0.91 & 0.94/0.94 & 0.89/0.89 & 0.89/0.89 \\
 & & Avg. & 0.89/0.89 & 0.89/0.89 & 0.88/0.88 & 0.9/0.9 & \colorbox{gray!30}{0.86/0.86} & 0.87/0.87 & 0.87/0.87 & 0.87/0.87 & 0.91/0.91 & 0.91/0.91 & 0.91/0.91 & 0.9/0.9 & 0.91/0.91 \\
\bottomrule
\end{tabular}
}
\end{subtable}

\begin{subtable}[t]{\textwidth}
\centering
\resizebox{\textwidth}{!}{%
\begin{tabular}{clccccc|cccc|ccc|cc}
\toprule
& & & \multicolumn{4}{c}{Baseline} & \multicolumn{4}{c}{Attention-wise} & \multicolumn{3}{c}{Neuron-wise} & \multicolumn{2}{c}{Uncertainty-wise} \\
Dataset & Model & Budget & Random & DeepGini & MaxP & Margin & KMAC & KVAC & KKAC & KSAC & IHNC & ITNC & FHNC & KMEC & KMLC \\
\midrule
\multirow{16}{*}{\vicuna} & \multirow{4}{*}{TruQA} & 5\% & 0.61/0.41 & 0.59/0.41 & 0.51/0.31 & 0.51/0.3 & 0.53/0.33 & 0.51/0.3 & 0.58/0.37 & 0.53/0.32 & 0.48/0.28 & 0.55/0.37 & 0.43/0.22 & 0.64/0.41 & 0.62/0.41  \\
&  & 10\% & 0.53/0.33 & 0.6/0.42 & 0.54/0.33 & 0.56/0.35 & 0.55/0.34 & 0.52/0.31 & 0.6/0.4 & 0.58/0.38 & 0.49/0.29 & 0.53/0.33 & 0.48/0.28 & 0.63/0.41 & 0.65/0.42  \\
&  & 15\% & 0.59/0.38 & 0.58/0.39 & 0.54/0.34 & 0.55/0.34 & 0.56/0.35 & 0.5/0.29 & 0.58/0.38 & 0.58/0.37 & 0.52/0.32 & 0.55/0.35 & 0.49/0.29 & 0.63/0.41 & 0.63/0.41  \\
 & & Avg. & 0.58/0.37 & 0.59/0.41 & 0.53/0.33 & 0.54/0.33 & 0.55/0.34 & 0.51/0.3 & 0.59/0.38 & 0.56/0.36 & 0.5/0.3 & 0.54/0.35 & \colorbox{gray!30}{0.47/0.26} & 0.64/0.42 & 0.64/0.42 \\
\cmidrule{3-16}
& \multirow{4}{*}{TriQA} & 5\% & 0.52/0.31 & 0.57/0.36 & 0.54/0.32 & 0.54/0.32 & 0.49/0.27 & 0.49/0.29 & 0.52/0.31 & 0.53/0.33 & 0.48/0.26 & 0.44/0.24 & 0.43/0.23 & 0.48/0.24 & 0.42/0.22 \\
&  & 10\% & 0.51/0.3 & 0.57/0.36 & 0.54/0.33 & 0.53/0.32 & 0.49/0.28 & 0.49/0.28 & 0.52/0.31 & 0.52/0.31 & 0.46/0.25 & 0.45/0.25 & 0.44/0.23 & 0.48/0.24 & 0.44/0.23  \\
&  & 15\% & 0.5/0.29 & 0.56/0.34 & 0.53/0.32 & 0.52/0.31 & 0.5/0.29 & 0.49/0.28 & 0.51/0.3 & 0.53/0.32 & 0.47/0.25 & 0.45/0.25 & 0.45/0.24 & 0.47/0.24 & 0.45/0.24 \\
 & & Avg. & 0.51/0.3 & 0.57/0.35 & 0.53/0.32 & 0.53/0.32 & 0.49/0.28 & 0.49/0.28 & 0.52/0.31 & 0.53/0.32 & 0.47/0.26 & 0.45/{0.24} & \colorbox{gray!30}{0.44/0.24} & 0.48/0.25 & {0.44/0.24} \\
  \cmidrule{3-16}
& \multirow{4}{*}{NQ} & 5\% & 0.55/0.34 & 0.56/0.35 & 0.58/0.37 & 0.58/0.37 & 0.59/0.38 & 0.59/0.39 & 0.57/0.37 & 0.55/0.36 & 0.49/0.28 & 0.49/0.27 & 0.48/0.26 & 0.67/0.44 & 0.68/0.44 \\
&  & 10\% & 0.57/0.36 & 0.53/0.31 & 0.58/0.37 & 0.58/0.37 & 0.58/0.37 & 0.6/0.39 & 0.58/0.37 & 0.56/0.35 & 0.5/0.28 & 0.5/0.28 & 0.48/0.27 & 0.66/0.43 & 0.66/0.43 \\
&  & 15\% & 0.56/0.36 & 0.54/0.33 & 0.57/0.36 & 0.57/0.36 & 0.58/0.37 & 0.59/0.39 & 0.57/0.37 & 0.56/0.35 & 0.51/0.3 & 0.52/0.31 & 0.5/0.28 & 0.67/0.43 & 0.68/0.43 \\
 & & Avg. & 0.56/0.35 & 0.54/0.33 & 0.58/0.37 & 0.57/0.36 & 0.58/0.37 & 0.6/0.39 & 0.58/0.37 & 0.56/0.35 & 0.5/0.29 & 0.5/0.29 & \colorbox{gray!30}{0.48/0.27} & 0.67/0.44 & 0.67/0.44 \\
 \cmidrule{3-16}
 & \multirow{4}{*}{RTP} & 5\% & 0.84/0.84 & 0.98/0.98 & 0.96/0.96 & 0.97/0.97 & 0.92/0.92 & 0.93/0.93 & 0.82/0.82 & 0.89/0.89 & 0.97/0.97 & 0.97/0.97 & 0.97/0.97 & 0.9/0.9 & 0.93/0.93  \\
&  & 10\% & 0.85/0.85 & 0.95/0.95 & 0.95/0.95 & 0.96/0.96 & 0.91/0.91 & 0.88/0.88 & 0.81/0.81 & 0.87/0.87 & 0.95/0.95 & 0.97/0.97 & 0.96/0.96 & 0.88/0.88 & 0.88/0.88  \\
&  & 15\% & 0.83/0.83 & 0.94/0.94 & 0.95/0.95 & 0.96/0.96 & 0.88/0.88 & 0.88/0.88 & 0.8/0.8 & 0.86/0.86 & 0.94/0.94 & 0.96/0.96 & 0.95/0.95 & 0.87/0.87 & 0.86/0.86  \\
 & & Avg. & 0.84/0.84 & 0.96/0.96 & 0.95/0.95 & 0.96/0.96 & 0.91/0.91 & 0.9/0.9 & \colorbox{gray!30}{0.81/0.81} & 0.88/0.88 & 0.96/0.96 & 0.96/0.96 & 0.96/0.96 & 0.9/0.9 & 0.89/0.89 \\
\bottomrule
\end{tabular}
}
\end{subtable}
\caption{RQ2 - Experimental Results for the performance of test prioritization for LLM. The results are shown as "Mean Absolute Error/Mean Squared Error", where a lower value indicates a better result. The best average results are highlighted in gray. 
(TruQA: \truthfulqa; TriQA: \triviaqa; NQ: \nqopen; RTP: \realtoxicityprompt)
}
\vspace{-10pt}
\label{table:RQ2_results}
\end{table*}

In this RQ, we compare the performance of test prioritization using the proposed criteria against baseline methods.
Tab.~\ref{table:RQ2_results} presents the results of test prioritization for LLMs.
We report the mean absolute error (MAE) and mean squared error (MSE), which are common metrics for evaluating prioritization performance.
Test prioritization was conducted with budgets of 5\

Our proposed metrics outperform the baseline method in test prioritization.
For example, in \llamaSevenB with \realtoxicityprompt, the MAE and MSE of the attention-wise metrics KKAC is the lowest (0.78 and 0.78 respectively); in \llamaSevenB with \triviaqa, the average MAE and MSE of the neuron-wise metrics ITNC is 0.49 and 0.29, respectively; in \vicuna with \nqopen, the average MAE and MSE of FHNC is 0.48 and 0.27.

Attention-wise criteria can provide effective prioritization, with
KMAC, KKAC, and KSAC emerging as effective metrics across multiple datasets and models, \eg, they achieve the lowest error rate seven times in dataset-model pairs.
They consistently provide lower MAE/MSE, suggesting that leveraging attention mechanisms for test prioritization is beneficial.
Neuron-wise criteria, also frequently achieve the best performance, \eg, FHNC delivers the best performance for \vicuna with \truthfulqa, \triviaqa, and \nqopen (0.47/0.26, 0.44/0.24, and 0.48/0.27).

Different models exhibit varying strengths depending on the criteria used, and the performance of these criteria also varies across different model-dataset pairs.
For instance, KSAC is good with \llamaTwoSevenB--\nqopen, while \vicuna benefits from KMEC.
Certain criteria, \eg, KSAC, consistently perform well across diverse datasets and models, indicating their robustness and adaptability.
Additionally, the criteria typically offer consistent performance under different budgets.

In practice, we recommend that users initially focus on the attention-wise and neuron-wise criteria for test prioritization, and consider other criteria for more sophisticated performance.
For instance, users might employ KSAC or FHNC during early development and then explore other attention-wise and neuron-wise metrics as the model moves toward deployment.

\begin{tcolorbox}[size=title, colback=white, breakable]
    {\textbf{Answer to RQ2:}
    The experimental result demonstrates that the proposed metrics are effective in providing test prioritization guidance and they typically have a better performance than the existing baseline methods.
    In practice, we recommend the user should first utilize the attention-wise and neuron-wise metrics and try other criteria for better performance.
    }
\end{tcolorbox}

\subsubsection{RQ3: \rqthree}
\label{subsubsec:rq3}

\begin{figure*}
\centering
\includegraphics[width=\linewidth]{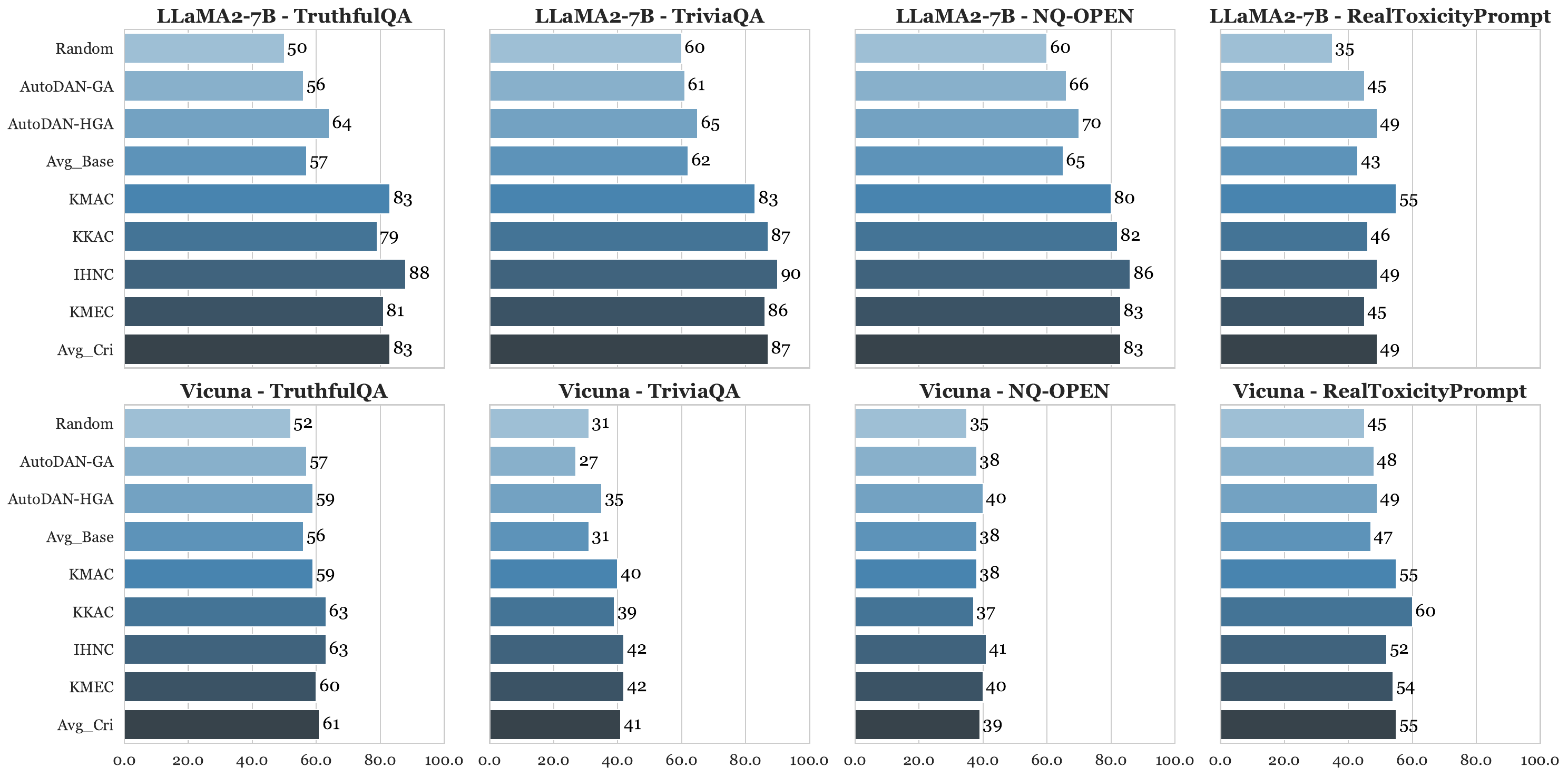}
\vspace{-12pt}
\caption{Test success rate (TSR) of coverage-guided testing. The x-axis is the TSR, and the y-axis is the testing method.
}
\vspace{-10pt}
\label{fig:cgt_tsr}
\end{figure*}

Fig.~\ref{fig:cgt_tsr} shows the test success rate of the baseline methods compared to the proposed coverage-guided testing.
The coverage-guided testing outperforms the baseline methods across all models and datasets.
The average TSR of the proposed criteria on \llamaTwoSevenB and \vicuna over all datasets are 76 and 49, respectively, while the ones of baseline methods are 57 and 43, respectively.
IHNC generally shows high scores across most models and datasets, \ie, it achieves the best of six trials over eight model-dataset pairs. 
For example, it gets a TSR of 90 for \llamaTwoSevenB--\triviaqa, and 86 for \llamaTwoSevenB--\nqopen.
This demonstrates the effectiveness of using the proposed criteria in coverage-guided testing.
For the baseline, the \autodan-HGA technique generally shows higher performance than the \autodan-GA and Random baselines across most models and datasets. 
For example, in the \llamaTwoSevenB model, the TSR of \autodan-HGA is 70 for \nqopen, compared to 66 for \autodan-GA and 60 for Random.

\begin{tcolorbox}[size=title, colback=white, breakable]
    {\textbf{Answer to RQ3:}
    Our coverage-guided testing is effective in guiding the testing procedure to find faults for diverse models and datasets.
    Criteria like IHNC generally yield higher test success rates and demonstrate their usefulness in practice.
    }
\end{tcolorbox}

\section{Related Work}
\label{sec:relatedwork}

\noindent\textbf{Testing Criteria of DNNs.}
Researchers have been actively exploring indicators to better represent DNN behavior patterns as criteria for guiding diverse quality assurance methods in DNN-centric systems, which can be categorized into two types: those relying on internal white-box analysis and those focusing on external black-box analysis. 
For the former, structural test coverage criteria have garnered significant attention~\cite{pei2017deepxplore, sun2018testing, xie2019deephunter, kim2019guiding}. 
They typically analyze neuron activation patterns, compute their distribution on both training and test data, and perform comparisons. 
This analysis can be extended to specific structures, such as RNNs, using internal-neuron-based stateful abstraction~\cite{du2019deepstellar, zhang2021decision}. 
For black-box analysis, both uncertainty-wise metrics (model-specific)~\cite{feng2020deepgini} and test data diversity metrics (model-agnostic)~\cite{ma2021test, aghababaeyan2023black} have proven effective in DNN testing. 
Different from most previous studies that focus on specific-purpose models (\eg, CNNs, RNNs) and classification tasks, we aim to explore LLM-specific criteria for general foundation models.

\noindent\textbf{Trustworthiness of LLMs.}
While LLMs demonstrate record-breaking performance on various downstream tasks, recent studies have highlighted concerns about their trustworthiness in practical applications.
Benchmarks~\cite{wang2023decodingtrust, zhang2023safetybench, sun2024trustllm} focusing on issues such as hallucination, safety, fairness, robustness, privacy, and machine ethics reveal that current LLMs are still far from perfect. In response to these challenges, efforts to improve LLM trustworthiness include designing better prompts for more appropriate responses~\cite{tan2023self, min2024beyond}, leveraging fine-tuning techniques for safety alignment~\cite{ouyang2022training, rafailov2024direct, ji2024beavertails}, and building safeguard systems to mitigate potential risks~\cite{markov2023holistic, inan2023llama}. Unlike these direct improvement approaches, our paper’s criteria focus on offline testing stages to identify potential trustworthiness issues. Additionally, the criteria studied in this paper can potentially enhance the effectiveness of the aforementioned improvement methods.

\section{Conclusion}
\label{sec:conclusion}

In this work, we propose \tool, a set of multi-level testing coverage criteria for LLMs.
\tool includes three types of criteria: attention-wise coverage, neuron-wise coverage, and uncertainty-wise coverage.
We apply these criteria to two application scenarios, \ie, test prioritization and coverage-guided testing, and the experimental results demonstrate their effectiveness and usefulness.
Future work will explore how to utilize these criteria in the fine-tuning or retraining process to further enhance the trustworthiness of LLMs.

\bibliographystyle{unsrtnat}
\bibliography{ref}

\end{document}